*A Technical Report on*
*Complex Systems*

**Jeffrey B. Smith**

**Advisor: Dr. Daniel Joyce**

**Completed: Fall 2002**

**Department of Computing Sciences**
**Villanova University**
**Villanova, Pennsylvania**
**U.S.A.**

# ABSTRACT

The study of Complex Systems is considered by many to be a new scientific field, and is distinguished by being a discipline that has applications within many separate areas of scientific study. The study of Neural Networks, Traffic Patterns, Artificial Intelligence, Social Systems, and many other scientific areas can all be considered to fall within the realm of Complex Systems, and can be studied from this new perspective. The advent of more capable computer systems has allowed these systems to be simulated and modeled with far greater ease, and new understanding of computer modeling approaches has allowed the fledgling science to be studied as never before.

The preliminary focus of this paper will be to provide a general overview of the science of Complex Systems, including terminology, definitions, history, and examples. I will attempt to look at some of the most important trends in different areas of research, and give a general overview of research methods that have been used in parallel with computer modeling. Also, I will further define the areas of the science that concern themselves with computer modeling and simulation, and I will attempt to make it clear why the science only came into its own when the proper modeling and simulation tools were finally available. In addition, although there seems to be general agreement between different authors and institutes regarding the generalities of the study, there are some differences in terminology and methodology. I have attempted in this paper to bring as many elements together as possible, as far as the scope of the subject is concerned, without losing focus by studying Complex System techniques that are bound to one particular area of scientific study, unless that area is that of computer modeling.

# 1. THE SCIENCE OF COMPLEXITY

The science of Complex Systems has been described by the New England Complex Systems Institute at MIT (NESCI) as "a new field of science studying how parts of a system give rise to the collective behaviors of the system, and how the system interacts with its environment. Complex systems have multiple interacting components whose collective behavior cannot be simply inferred from the behavior of components. The recognition that understanding the parts cannot explain collective behavior has led to various new concepts and methodologies that are affecting all fields of science and engineering, and are being applied to technology, business and even social policy" (NECSI website). This collective approach is a relatively new strategy within science in general, as research techniques in the past have typically been reductionist, or geared toward breaking objects apart in order to understand them. Scientists that study Complex Systems maintain that these systems have key components and comprehensible properties in common, even though these systems exist in separate branches of science (Lamper, Website).

A fundamental reason why this field of study has only recently been explored has to do with the availability of proper tools to study and simulate these systems. The human mind has difficulty keeping track of "many and arbitrary interacting objects or events – we can typically remember seven independent pieces of information at once" (Bar-Yam, xii). The computer is useful as a tool in areas where the human brain is not as useful: it is able to remember and precisely simulate the interaction of a large number of elements. Just as it would be impossible to underestimate the role of the microscope when studying Biology, the role of the computer has been essential for the study of Complex Systems. The science has been around for years, as evidenced by the work of Vemuri in his book, *Modeling of Complex Systems*, from 1978. However, the science has attempted to use other tools like mathematics, number theory, statistics, to understand different varieties of Complex Systems, and it is not until the computer became a viable research tool that a more complete study of these systems was possible (Bar-Yam, xiii).

Some of the best-known examples of Complex Systems are systems we interact with everyday: The weather, the stock market, traffic on the road to work, etc. The vast complexities of humanities' interaction with the world, of the individual to society, the permutations of the brain, and the interactions and behavior of an ant colony are all examples of systems with different levels of complexity.

## 1.1 What is a Complex System?

The NECSI website describes a Complex System as having "multiple interacting components whose collective behavior cannot be simply inferred from the behavior of components. The recognition that understanding the parts cannot explain collective behavior has led to various new concepts and methodologies that are affecting all fields of science and engineering, and are being applied to technology, business and even social policy" (NECSI website). A Complex System is a system which has these properties:

- It has more than a few and less than too many parts/elements.
- These parts/elements are heterogeneous, and must interact in a non-linear fashion.

- It has a definite purpose, objective, function.
- It does not operate in equilibrium i.e. it is adaptive, dynamic, and always changing.
- It has collective behavior that cannot be inferred from its parts/elements (Cilliers, 5).

This definition is by no means complete. The field of study is new enough so that scientists will expound upon the properties of a Complex System, but are unwilling to commit to very many specific laws that define these systems. The rules above are ones that were universally described in the research materials used, but for further clarification, Paul Cilliers' complete list of properties are included in Appendix I of this study (CSCS Website). What they essentially state is that a Complex System is a system that cannot easily be explained in a mathematical fashion, and contains too many elements for a formal description of behavior to be presented.

### *1.1.1 Definitions*

What do we mean when we discuss a Complex System, or a *highly non-linear complex heterogeneous system*? The first important distinctions to make are between *simple*, *complex*, and *complicated* systems. These distinctions are often a function of our *distance* from the system, because a hidden complexity in a system can be masked by a simple presentation, and a complex system can exist "below" the level in which we are interested. For example, a complex traffic pattern might be a system we are interested in studying, while the occupants of the vehicles that compose the pattern are not interesting to us. We may also be interested in the flight patterns of a large flock of Canadian Geese, but we may not be interested in the color patterns of their feathers. Our frame of reference, known as our *frame* in Complex System language, is the level of complexity we are currently interested in studying.

A *simple* system is one that can typically be understood by explaining the properties of a single element/part, or the interactions of a few elements/parts. Simple systems have long been the domain of traditional scientific investigation, and explain the properties of mysteries like the motion of the Earth around the Sun, or a nuclear chain reaction.

A *complicated* system is actually a simple system, but it is disguised as something more complex. The classic example of this is the Brownian motion of gas in a vacuum. Even though many molecules make up the gas, the overall properties of the gas are explained by conventional laws of Thermodynamics. It is a system with many moving parts, but little complexity (Cilliers, 3).

The fundamental rule that distinguishes these systems is called *emergence*. An *emergent* property is a property that exists in a group of elements in a Complex System, but does not exist in each individual element. The previous example of a gas under pressure describes this concept well: The gas has properties like pressure and temperature, which are properties that the separate molecules do not possess. Complex Systems possess different degrees of emergence, both *local* and *global*. The gas properties of pressure and temperature are examples of *local emergence*, with just a small number of elements exhibiting the same properties as a large number of elements in a similar situation. A Complex System is distinguished by its display of *global emergence*, or properties that are present only in the entire system, and not in the elements or groups of elements. The classic example is the human brain, in which a neuron or a group of neurons can "remember" specific memory patterns. If neurons are removed from this group, the whole system may lose its ability to remember any memory patterns. This is not true with a gas, in which elements are interchangeable and removable, and the removal of a part will not change the attributes of the whole (Bar-Yam, 10).

It is not yet clear how dramatic this distinction is, and just as it may be gradual shift from local to global emergence, it may also be a slow change from a complicated to a complex system. An analysis of traffic patterns in New York City street grid may only be a complicated system, but the addition of a Broadway street that breaks up the grid pattern and a Knicks game at Madison Square Garden may make the traffic pattern a Complex System. A true Complex System may possess both global *emergent complexity* and global *emergent simplicity*, which means that collections of simple objects can possess complex emergent properties and collections of complex objects can possess simple emergent properties (Bar-Yam, 293).

The fundamental building block in a Complex System is known as an *agent*. Previously this has been referred to as an *element* or *part*, but a Complex System usually has elements that can be distinguished by having specific roles and attributes, and so *agent* is more appropriate. Agents can be very simple constructs, like the molecules in Brownian motion, but more typically has a complicated *strategy* and many *attributes*, with varying rules of both *linear* and *non-linear* interaction (Axelrod, 4). This generally means that for any given action upon it, an agent will exhibit behavior, and sometimes this behavior will be out of proportion to the action. For example, a free gas molecule will always exhibit a linear response to an action upon it. In a Complex System, it is possible that an agent will exhibit no reaction when action is taken on it the first time, and then exhibit a reaction that is twice the magnitude of the action the second time it is acted upon (NECSI website).

The agents can be categorized into specific *types*. For example, in the free gas molecule scenario, some of the molecules could be Nitrogen, and some water vapor, and they may act differently in response to a stimulus. In a more correct example, when analyzing traffic patterns in New York City, some of the agents could be of the "truck" type, and some of the "automobile" type. All of these agents are gathered together to form a *population* (Axelrod, 4).

The entire Complex System will be gathered together in an *environment*. This is entirely dependent upon how the system is *framed* by the *observer*, as mentioned previously. The observer is generally outside the system, just as the environment exists around the system, but the possibility exists that either could exert influence on the system. The Heisenberg Uncertainty Principle is a useful analogy that serves to illustrate this point: By attempting to measure the momentum of an electron, it is necessary to bounce photons off of it, which will change the momentum of the electron and give the observer incorrect results, and also result in a change of state in the electron. The influence that is exerted and received from a Complex System is one of the reasons they are innately difficult to study. In another field of study, an anthropologist studying a hunter-gatherer tribe in South America would find it difficult to study the tribe without becoming a part of the tribe. As a result, Bar-Yam of NECSI presents the following definition: "An observer is a system which, through interactions, retains a representation of another system (the observed system) within it" (NECSI website).

*1.1.2 Concepts*

Some explanation has been given as to the interactions that can be expected to occur in a Complex System, but more definition is needed. Specifically, how can these interactions be characterized, and can any fundamental rules of behavior be determined? The best way to explain this is to classify the different kinds of Complex Systems.

*1.1.2.1 Non-Adaptive Complex Systems*

The Non-Adaptive Complex System (NACS) is a system composed of agents without a choice. They do not change their rules of behavior during the operation of the system. While the objects interact with each other, they follow predictive rules, without exception. They follow the same rules under all conditions of interaction (Casti, website). The "Game of Life" is a good example of such a system. The various agents are placed in a starting position on the game-board, and the simulation starts. Each event after this point is determined by the rules, which decide whether an agent "survives" into the next round, or is eliminated. The agents do not adapt, and the success of the system is determined by the starting locations of the agents. The rules for this game are explained below, in Section 1.1.4.

*1.1.2.2 Complex Adaptive Systems*

The Complex Adaptive System (CAS) is a collection of agents which are in interaction with each other and that have a choice of rule systems which are followed during the interaction (Casti, website). Naturally, a CAS can also contain aspects of a NACS, and can contain agents that do not change their behavior. Similarly, an aspect of a particular CAS may be that the agents only change their behavior at specified moments. What is important is that the possibility exists that the agents will display *adaptive* behavior. Many researchers maintain that it is this adaptability that makes the system complex, but the field is divided on this issue (Axelrod, 9).

*1.1.2.3 Interactions*

Much of the terminology for a Complex Adaptive System has already been introduced, but there are a few key concepts introduced by Robert Axelrod that characterize the interaction of the agents in such a system:

The system is generally composed of a population or a multiple population of agents of varying types, some adaptive, some non-adaptive, all interacting in a *concurrent*, *co-evolutionary* fashion. The co-evolutionary or co-adaptive process follows a model that is already known in evolution and genetic programming, in which mutation, crossover, extinction, birth, and recombination occurs between populations of agents (see Section 1.1.3.4). Essentially this means that all agents are interacting and surviving based upon their own criteria for success, and slowly evolving based on either learned responses (in the field of Sociology or Anthropology) or based on reproduction (as in the case of genetic algorithms). In fact, it is not always clear if the system is adaptive based on the individual agents or the emergent properties of the system. For example, in the case of a Neural Network, it might be understood that the system is adaptive, and not the agents (Axelrod, 40). The other aspect is that the agents are doing all of this concurrently, which simply stated means "simultaneous occurrence" (Ghosh, 23).

Paul Cilliers maintains that the set of rules that govern a Complex System can be *abstract*, or agent and type based, or they can be *centralized*, or both. This concept is not a surprise, especially within the study of Politics or Economics. It's important to note that this control can occur at any level in a Complex System (Cilliers, 3).

There are many other specific descriptions of agent interaction, but most are specific to the type of Complex System being considered, and would be too numerous to detail here. Many of the terminology here has been used before to describe Complex Systems in a multitude of forms, and it is perhaps for this reason that the field has not agreed on a set standard.

### 1.1.3  Methods

#### 1.1.3.1  Computer Modeling and Simulation

This is probably the most important of the methods used in studying Complex Systems, and because it is the main focus of this paper, it will be described in detail in Chapter 3.

#### 1.1.3.2  Measuring Complexity

It is important to note that not all researchers in the field of complex systems are in agreement on the best way to measure complexity (Axelrod, 16). Nevertheless, Bar-Yam references several Computer Science methods that can be used in measuring complexity, and provides some explanation on how they might be applied to Complex Systems. He postulates that "the complexity of a system is the amount of information needed to describe it" (Bar-Yam, 12). From this declaration, he bases his measurement of Complex System complexity upon methods that are well-known in the realm of Computer Science: "If we have a system that could have many possible states, but we would like to specify which state it is actually in, then the number of binary digits (bits) we need to specify this particular state is related to the number of states that are possible. If we call the number of states $\Omega$ then the number of bits of information is needed is:

**$I=\log_2(\Omega)$**"

(Bar-Yam, 13). This is familiar to Computer Scientists as the "Big-O" method of measuring the complexity of an algorithm. Bar-Yam goes on to explain that the number of possible states $N$ of a system is related to the number of bits $B$ by $B = 2^N$. It is important to remember that a Complex System is often composed of many sub-systems that contain their own level of complexity, and the complexity of the whole may be the sum and/or the product of its parts, if the sub-systems interact. Because this level of complexity is quite difficult to describe, Bar-Yam maintains that it is essential to define a complexity profile that is a function of the scale of observation (frame). This is what allows a Complex System to be studied from a human perspective. Also, the complexity of the system must be a monotonically falling function of the scale. This makes intuitive sense, because the level of detail needed to describe the system at a higher level must always be less than at a lower level, and the behavior of the upper level may often be described in terms of the emergent properties of the lower level (Bar-Yam, 14).

*1.1.3.3 Chaos Theory*

Another important concept that needs to be addressed is chaos. The scientific and mathematical study of Chaos Theory contains many overlaps with the study of Complex Systems, but with differences related to method: Chaos Theory can be used to study Complex Systems, but is not restricted to the study of these systems. Chaos Theory "deals with deterministic systems whose trajectories diverge exponentially over time" (Bar Yam, NECSI website). It has been used to study Complex Systems, because these systems can be generally defined as a "deterministic system that is difficult to predict". However, "chaos deals with situations such as turbulence that rapidly become highly disordered and unmanageable. On the other hand, complexity deals with systems composed of many interacting agents" (Axelrod, xv). The point being that Chaos Theory is one of many tools and methods that can be applied to the study of Complex Systems, but is not specifically devoted to the way these systems are designed, developed, studied, and modeled. That being stated, the famous example of the "Butterfly Effect" in a chaotic system is an example of an agent (a butterfly) evoking a non-linear response (the storm in New England) within a Complex System (Global Weather System).

*1.1.3.4 Genetic programming*

Genetic Programming, a form of *automatic programming*, is related to the field of Complex Systems in the sense that it is a powerful technique that can be used to harness complexity, and solve complex problems. Essentially, populations of algorithms are created in a similar state, and each attempt to find a solution to a problem. Each are selected or discarded based on the success of their solution to the problem. A technique called *mutation* (roughly analogous to the same process in Evolutionary theory), *crossover*, and *recombination* occurs after each selection process, in which new agents are generated and allowed to attempt to solve the problem. After a succession of these "reproductive" cycles, a solution is found, and often a better solution than a laboring engineer could come up with. It is in this fashion that programmers can harness complexity to solve difficult problems. In a sense, the genetic algorithm population represents a Complex Adaptive System, with the solution to the problem being the most important emergent property, and the individual algorithms displaying adaptive behavior (Axelrod, 10).

*1.1.3.5 Other Methods*

There are a great many other methods that are used within the study of Complex Systems. This paper looks specifically at Computer Modeling of Complex Systems, and therefore does not focus on methods and approaches that lie outside the realm of Computer Science, but Yaneer Bar-Yam outlines a number of approaches in his book *Dynamics of Complex Systems*, and the most important of these are listed below:

- *Statistical Methods*: Statistical methods have been used in all sciences, particularly in the Social and Biologic Sciences for the capturing of experimental results within large study populations. They have also been used to study any scientific scenario that utilizes entity or agent populations (Cilliers, 10).

- *Stochastic Iterative Maps*: These are systems in which the outcome of a particular decision point is probabilistic and not deterministic. It is a mathematical tool that can be used in combination with computer modeling (Bar-Yam, 43).

- *Neural Network methods*: A series of methods and model used for understanding the mind can offer insight into other Complex Systems as well. Concepts like the *connectionist approach*; used to describe the intensity of connections between neurons; is often used as a model to help understand other systems with agents that cannot be easily disseminated into a discrete set of rule-based interactions (Cilliers, 25).

- *Mathematical and Analytical Models*: This has been the most prevalent approach to Complex Systems before the advent of computer modeling. Analytical Models are developed to capture system behavior through exact equations, which are then solved using mathematical techniques (Ghosh, 1).

- *The Theory of Computation*: This uses familiar concepts like the Turing machine to measure the complexity in a system. It may also be used to analyze agent interactions, which can be very complex (Bar-Yam, 243).

- *Thermodynamics and Statistical Mechanics*: A Physics approach to study the action of a population of many agents, much like one might study gas molecules in a container. This uses statistical methods to improve results (Bar-Yam, 59).

### 1.1.4  Examples

#### 1.1.4.1  The Game of Life

One of the most interesting introductory examples of a Complex System is John Conway's "Game of Life". This example is interesting because it is a very clear demonstration of a non-adaptive Complex System, in which very simple agents display a clear example of emergent complexity. Paul Callahan of Princeton University explains this game best as follows: "The Game of Life (or simply Life) is not a game in the conventional sense. There are no players, and no winning or losing. Once the "pieces" are placed in the starting position, the rules determine everything that happens later. Nevertheless, Life is full of surprises! In most cases, it is impossible to look at a starting position (or pattern) and see what will happen in the future. The only way to find out is to follow the rules of the game.

Life is played on a grid of square cells--like a chess board but extending infinitely in every direction. A cell can be live or dead. A live cell is shown by putting a marker on its square. A dead cell is shown by leaving the square empty. Each cell in the grid has a neighborhood consisting of the eight cells in every direction including diagonals. To apply one step of the rules, we count the number of live neighbors for each cell. What happens next depends on this number.

- A dead cell with exactly three live neighbors becomes a live cell (birth).

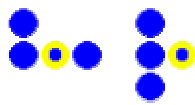

- A live cell with two or three live neighbors stays alive (survival).

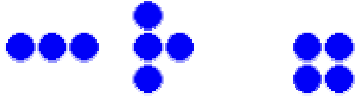

- In all other cases, a cell dies or remains dead (overcrowding or loneliness).

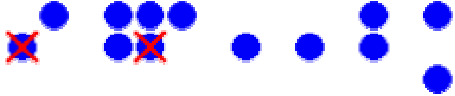

Life is just one example of a cellular automaton, which is any system in which rules are applied to cells and their neighbors in a regular grid. There has been much recent interest in cellular automata, a field of mathematical research. Life is one of the simplest cellular automata to have been studied, but many others have been invented, often to simulate systems in the real world.

In addition to the original rules, Life can be played on other kinds of grids with more complex patterns. There are rules for playing on hexagons arranged in a honeycomb pattern, and games where cells can have more than two states (imagine live cells with different colors).

Life is one of the simplest examples of what is sometimes called *emergent complexity* or *self-organizing systems*. This subject area has captured the attention of scientists and mathematicians in diverse fields. It is the study of how elaborate patterns and behaviors can emerge from very simple rules. It helps us understand, for example, how the petals on a rose or the stripes on a zebra can arise from a tissue of living cells growing together. It can even help us understand the diversity of life that has evolved on earth" (Callahan, Website).

*1.1.4.2  Jane Jacobs: The Death and Life of Great American Cities*

Author Jane Jacobs has written a number of books that look at Complex Systems from a sociological and analytical perspective. Books like *Systems for Survival* and *Cities and the Wealth of Nations* all focus on the connectivist aspects of economies and societies. The example of a Complex System that is perhaps easiest to see is her dissemination of New York City in the *Death and Life of Great American Cities*, which takes the city apart level by level and block by block, so that the connections between street life, safety, commerce, and government are easy to see. The ultimate goal in her writing is to show the urban designers of the late 60's that the model being used to represent the city is inappropriate and oversimplified, but the byproduct of her work is a sharply analyzed mental model of the way agents in a city interact. Jacobs introduces a number agent types, the emergent traits of city blocks, and the interconnections that make the city possible. Her work has helped lead to a reanalysis of the way complex systems like the City have been studied.

Since this time, there have been many attempts to model the socio-economic forces within cities, to the point in which this category of modeling has become a separate branch of sociology called *Artificial Life*, which uses agent-based modeling to examine humanity in general (Jacobs, Introduction).

## 1.2 A Brief History

Some of the first concepts in the study of Complex Systems were introduced by Adam Smith in 1776, in which he described the emergent properties of the stock market. Naturally, Charles Darwin also introduced many key concepts, including some terminology. Also responsible was Henri Poincare and his work with Non-linear systems, which illustrated apparent casual systems exhibiting non-deterministic and complex behavior (Antti-Poika, Website). However, the study of Complex Systems really began to catch on in the latter half of the 20th century, and most notably in several different areas of scientific study. These disciplines included:

- Condensed matter physics, with its concern with nonlinear interactions among the spins of many particles and the retention of patterns that reduce "frustration" between the elements of a system.

- Evolutionary Biology, with its concern for populations with "gene pools" of strategies that evolve through selective reproduction with variation.

- Evolutionary computation (Genetic Programming), the branch of Computer Science that is inspired by Evolutionary Biology (see Section 1.1.3.4).

- Social Science modeling of heterogeneous populations of people who interact and thereby exhibit adaptive behavior.

- Cellular Automata, which is populations of very simple, locally connected computing elements (see Section 1.1.4.1).

- Artificial Life, the study of many different systems, usually implemented as computer simulation agents that exhibit lifelike properties (similar to Computer Simulation and Modeling).

- Mathematical Theories to formalize the measurement of complexity within a system, which can include Complexity Theory and Chaos Theory.

- Artificial Intelligence, which is the study of reproducing intelligence and consciousness with a Computer program.

(Axelrod, 17).

It would be a lesson in 20th century scientific study to try to show all the advances that led to the development of these separate areas of study, and to show how Complex Systems arose from this. What's important is that advances in these separate areas of study all led to the consideration of these systems as having properties in common that might aid in understanding and researching them. This "holistic" approach was championed at several universities at around the same time, the most notable being the New England Complex Systems Institute at MIT, the Center for the Study of Complex Systems at the University of Michigan, the COSA Research Consortium, and the Santa Fe Institute. Intersecting with these organizations is the *Journal of Artificial Societies and Social Simulation*, a variety of research centers dedicated to Computer Modeling and Simulation, and many others. It is because of the fact that so many sciences needed to recognize the universality of Complex Systems that it took so long for this recognition to occur.

## 2. BIBLIOGRAPHY

1. Antti-Poika, Teemu. *Complex Systems and their Interdisciplinary Applications in Science*, COSA Research Consortium Website, 2000. http://www.math.jyu.fi/research/cosa/
2. Axelrod, Robert and Michael D. Cohen. *Harnessing Complexity: Organizational Implications of a Scientific Frontier*, The Free Press, 1999.
3. Axelrod, Robert. *The Complexity of Cooperation: Agent-Based Models of Competition and Collaboration*, The Princeton University Press, 1998. http://pscs.physics.lsa.umich.edu/Software/ComplexCoop.html
4. Bar-Yam, Yaneer. *Dynamics of Complex Systems*, Addison-Wesley, 1997.
5. Ben-Dov, Yoav. *Complexity and Emergence*, Interview with Chris Langton, Sante-Fe Institute, 2002. www.bendov.info/eng/langton.htm
6. Callahan, Paul. *What is the Game of Life?*, Math.com Website, 2000. http://www.math.com/students/wonders/life/life.html
7. Casti, John. *The Science and Surprise of 'Would-be' Artificial Worlds: The Collaborative Process of Developing an Industry-wide Model*, The LSE Strategy & Complexity Seminar, 1998. http://bprc.warwick.ac.uk/Casti.html
8. Cilliers, Paul. *Complexity and Postmodernism: Understanding Complex Systems*, Routledge, 1998.
9. *Complex Systems Virtual Library*, Charles Stuart University, 2000. http://lorenz.mur.csu.edu.au/vl_complex/topics.html
10. *CSCS: Center for the Study of Complex Systems*, The University of Michigan, 2002. http://www.pscs.umich.edu/complexity.html
11. Dooley, Kevin. *Complex Adaptive Systems: A Nominal Definition*, Arizona State University Website, 1996. http://www.eas.asu.edu/~kdooley/casopdef.html
12. Epstein, Joshua and Robert Axtell. *Growing Artificial Societies: Social Science from the Bottom Up*, The Brookings Institution Press, 1996. http://www.brookings.edu/dybdocroot/Sugarscape/Default.htm
13. Fogel, David B. *Evolutionary Computation: Toward a New Philosophy of Machine Intelligence*, IEEE Press, 2000.
14. Gaylord, Richard J. and Louis J. D'Andria. *Simulating Society: A Mathematica Toolkit for Modeling Socioeconomic Behavior*, Springer-Verlag New York, Inc. 1998.
15. Ghosh, Sumit, and Tony Lee. *Modeling and Asynchronous Distributed Simulation: Analyzing Complex Systems*, IEEE Press, 2000.
16. Gilbert, Nigel. *JASSS: The Journal of Artificial Societies and Social Simulation*, University of Surrey, 2002. http://jasss.soc.surrey.ac.uk/JASSS.html
17. Holmevik, Jan Rune. *The History of Simula*, University of Trondheim, 1995. http://java.sun.com/people/jag/SimulaHistory.html

18. *IDSIA, A Collection of Modelling and Simulation Resources on the Internet*, University of Lugano and SUPSI, 2001. http://www.idsia.ch/~andrea/simtools.html

19. Jacobs, Jane. *The Death and Life of Great American Cities*, Vintage Books, 1961.

20. Koeppel, Dan. "Massive Attack", *Popular Science*, December 2002, Volume 261, Number 6.

21. Koza, John et al. *www.genetic programming.org*, Stanford University, 2000. http://www.genetic-programming.org/

22. Lamper, David, and Neil F. Johnson *The Science of Complexity*, Dr. Dobbs Journal, 2002. http://www.ddj.com/documents/s=7570/ddj0210a/0210a.htm

23. *Mathematica: The Way the World Calculates*, Wolfram Research, Inc., 2002. http://www.wolfram.com/products/mathematica/index.html

24. *NECSI: New England Complex Systems Institute*, 2002. http://necsi.org/

25. *Staelin, Charles P. jSIMPLEBUG: A Swarm Tutorial for Java*, Smith College, 2000. http://eco83.econ.unito.it/swarm/materiale/jtutorial/jSimpleBug.pdf

26. *The Swarm Development Group*, Sante Fe Institute, 2002. http://www.swarm.org/

27. Vemuri, V. *Modeling of Complex Systems: An Introduction*, Academic Press, Inc. 1978.

28. Whicker, Marcia Lynn, and Lee Sigelman. *Computer Simulation Applications: An Introduction*, Sage Publications, Inc., 1991.